\documentstyle{article}\begin{document}
\title{Thermodynamics of diffusive DM/DE systems }
\author{ Z. Haba\\
Institute of Theoretical Physics, University of Wroclaw,\\ 50-204
Wroclaw, Plac Maxa Borna 9, Poland\\
email:zhab@ift.uni.wroc.pl}\maketitle
\begin{abstract}We discuss the energy density, temperature and entropy
of dark matter (DM) and dark energy (DE) as  functions of the
scale factor $a$ in an expanding universe. In a model of
non-interacting dark components we repeat a derivation  from
thermodynamics of the well-known relations between the energy
density, entropy and temperature. In particular, the entropy is
constant as a consequence of the energy conservation. We consider
a model of a DM/DE interaction where the DM energy density
increase is proportional to the particle density. In such a model
the dependence of the energy density and the temperature on the
scale factor $a$ is substantially modified.  We discuss (as a
realization of the model)  DM which consists of relativistic
particles diffusing in an environment of DE. The energy gained by
the dark matter comes from a
 cosmological fluid with a negative pressure.
 We define the entropy and free energy of such a non-equilibrium
 system. We show that during the universe evolution the entropy of DM is increasing
whereas the entropy of DE is decreasing. The total entropy can
increase (in spite of the energy conservation) as the DM and DE
temperatures are different. We discuss non-equilibrium
thermodynamics on the basis of the notion of the relative entropy.
   \end{abstract}

\section{Introduction}
The universe evolution is determined by Einstein equations with an
energy-momentum tensor on their rhs. The energy-momentum is to
contain an information about a large system of \textquoteleft
particles\textquoteright ( galaxies, interstellar gas, dark matter
etc.). A deterministic description of such a system is
unrealistic. The successful $\Lambda$CDM model \cite{bicep}
determines the global evolution of the universe. In order to go
into its more detailed structure  we must construct particular
models of the energy-momentum (cosmological fluids, inflaton
fields). Simple models involve non-interacting components of the
energy-momentum. However, if we wish to explain a relation between
the components and their evolution in time we need to take into
account an interaction between components. Some basic properties
of large systems are a consequence of their statistical  and
thermodynamic description (as their macroscopic version). The
thermodynamic point of view has been developed from the beginning
of cosmology \cite{tolman}. It concerned non-interacting fluids.
 In this paper we discuss some
consequences of an interaction.Einstein equations have the
energy-momentum tensor on their rhs which is conserved. As a
consequence, in a homogeneous universe the  entropy of matter
consisting of non-interacting components is also conserved.
However, if the energy momentum consists of components which
exchange energy then the corresponding entropies of the subsystems
change in time. We consider a system consisting of radiation, \textquoteleft
baryonic\textquoteright 
 matter, dark matter
(DM)and dark energy(DE). The energy momentum of \textquoteleft
baryonic\textquoteright  matter and radiation is conserved. The
energy-momentum of the dark matter is not conserved and its
divergence has the opposite sign with respect of the divergence of
the energy momentum of the dark energy. We propose a particular
model of the energy-momentum non-conservation. The change of the
energy-density is proportional to the particle density. In Sec.3
we study some thermodynamic consequences of this assumption. In
Sec.4, based on our earlier papers \cite{habacqg,hss} we
introduce a particular realization of this energy-momentum
non-conservation resulting from a diffusion of DM in a fluid of
DE. In Sec.2 we review some old derivations of the relations
between energy density, temperature and entropy ( the derivation
follows the text-book of Planck \cite{planck}) as a consequence of
the first law of thermodynamics or the energy-momentum
conservation. In Sec.3 we show how these relations are changed if
we have components of the energy-momentum which are not conserved.
We derive a formula for the energy density. On a basis of general
thermodynamic principles we suggest a formula for a dependence of
the temperature on the scale factor $a$ of an expanding universe.
Then, using the Clausius-Caratheodory concept of the temperature
as the integrating factor for the heat we can calculate the
entropy of each component of the cosmological fluid. The entropy
of the universe has been discussed in many papers
\cite{tolman,weinberg,frautschi,landsberg,gron,wallace,egan,frampton}.
The discussion concerned mainly non-interacting fluids. The large
value of the estimated entropy of the universe poses a serious
problem of its origin. It clearly contradicts the mechanical
\textquoteleft adiabatic\textquoteright  evolution of the universe
based on the $\Lambda$CDM model.  In our model the total entropy
of the fluids is increasing in spite of the energy conservation
(in contradistinction to non-interacting fluids). In Sec.5 we
discuss thermodynamics of diffusing systems using the notion of
relative entropy \cite{lebowitz,risken}.
 It allows to follow the relation between entropy, energy and the free energy.
 There are many computer simulations of structure formation in
which the entropy of the components of the cosmic fluid gives an
important information about the future evolution of the system and
exchange of heat inside the system \cite{basu,got}. In this paper,
in a particular model, we study the temperature of the components
and an evolution of their entropy. In sec.6 we discuss also the
generalized second law of thermodynamics
\cite{bekenstein,gibbons,davis} (taking into account the entropy
of the horizon). The entropy of the horizon is the dominating
contribution to the universe entropy \cite{egan,frampton} , but it
does not change the conclusions about the entropy increase in the
models of diffusive interactions. In the Appendix we show how to
extend our results to an arbitrary DM/DE interaction.

\section{Thermodynamics
of non-interacting DM/DE systems} As an introduction, let us begin
with a simplified description of equilibrium thermodynamics and
its extension to an expanding universe (for a more general
approach see \cite{maartens}). Let us consider the first law of
thermodynamics in the form\begin{equation} TdS=dE+pdV,
\end{equation}
where $S$ is the entropy, $T$ temperature, $E$   energy and $p$ is
the pressure. We define the energy density $\rho$ by
\begin{equation} E=\rho V,
\end{equation}where $V$ is the volume.
Let us assume the equation of state
\begin{equation}
p=w\rho,
\end{equation}
where $w$ is a constant. Then, treating $S(T,V)$ as a function of
$V$ and $T$ we have
\begin{equation}
\Big(\frac{\partial S}{\partial
T}\Big)_{V}=\frac{V}{T}\frac{d\rho}{dT}.
\end{equation}
\begin{equation}
\Big(\frac{\partial S}{\partial V}\Big)_{T}=(1+w)\frac{\rho}{T}.
\end{equation}
Comparing second order partial derivatives we get
\begin{equation}
\Big(\frac{\partial\rho}{\partial
T}\Big)_{V}=w^{-1}(1+w)\frac{\rho}{T}.
\end{equation}
Hence,
\begin{equation}
\rho(T)=BT^{\frac{1+w}{w}}
\end{equation}
and from eq.(5)
\begin{equation}
S=(1+w)BVT^{\frac{1}{w}}.
\end{equation}
with a certain constant $B$. As a consequence, for a negative
$w>-1$ the energy density $\rho$ as well as the entropy  $S$ are
decreasing functions of temperature. The derivation of eqs.(4)-(8)
for $w=\frac{1}{3}$ (corresponding to massless particles) appeared
already in the text-book of Planck (\cite{planck},eqs.(74)-(77))
leading to the Stefan-Boltzmann law $\rho\simeq T^{4}$.
Eqs.(7)-(8) make no sense for $w=0$. For a small $w$ eq.(7) should
be understood in the sense $T\simeq w\ln \rho$ and eq.(8) as
$T\simeq w\ln (\frac{S}{V})$ meaning that in the limit
$w\rightarrow 0$ the temperature depends neither on density nor on
the entropy.
 The formulas can still apply in an expanding universe.If we demand that in an expanding flat universe with $V=V_{0}a^{3}$ (where $a$ is the expansion scale factor)
 the entropy is constant (an adiabatic expansion) then from eq.(8)
($a(0)=1$ )\begin{equation} T_{a}=T_{0}a^{-3w}.
\end{equation}
We obtain from eq.(9) the well-known $a^{-1}$ decrease of the
temperature for radiation (with $w=\frac{1}{3}$ ).

Let us show that eq.(9) is in agreement with the energy-momentum
conservation
\begin{equation}
(T^{\mu\nu})_{;\mu}=0
\end{equation} and its representation as an ideal fluid

\begin{equation}
T^{\mu\nu}=(\rho+p)u^{\mu}u^{\nu}-pg^{\mu\nu} \end{equation}
 with a four-velocity $u^{\mu}u_{\mu}=1$. After choosing
the comoving frame $u=(1,0)$  the conservation of the
energy-momentum tensor takes the form\begin{equation}
\partial_{t}\rho+3H(\rho+p)=0,
\end{equation} where $H=a^{-1}\partial_{t}a  $.Inserting $V=V_{0}a^{3}$ in eq.(1) we derive from
eq.(12)
\begin{equation}
dS=0
\end{equation}
confirming that the  entropy is a constant and as a consequence
confirming also eq.(9). Moreover, we can solve eq.(12) with the
result
\begin{equation}
\rho=\rho_{0}a^{-3(1+w)}.
\end{equation}
Comparing with eq.(7) we get again the formula (9)(which was
derived from eq.(13) as resulting from the energy density
expansion law). So, the temperature law (7) follows either from
the conservation of energy or from the conservation of entropy.

If we have a non-interacting DM/DE system of ideal fluids then the
entropy of each component is constant. From now on $w$ will
describe the equation of state (3) for dark energy and the
corresponding equation of state  for the dark matter  will be
written as $\tilde{p}=\tilde{w}\tilde{\rho}$.  With $\tilde{w}>0$
and $w<0$ the density of DM is decreasing in time faster then the
density of DE (the density of the phantom DE with $w<-1$ is
increasing). The temperature of DM is decreasing whereas the
temperature of DE is increasing. Hence, there cannot be any
equilibrium between DM and DE. We have also the coincidence
problem that the densities of DM and DE are of the same order only
at the present time.

It is known that thermodynamics of non-ideal or inhomogeneous
fluids substantially modifies the temperature and energy density
evolutions \cite{weinberg,lima1,lima2,lima3}. We may expect that
an interaction of fluids will also have a profound effect of the
thermodynamics of components. This could be seen by a division of
a non-homogeneous fluid into homogeneous interacting parts.

\section{Thermodynamics of the DM/DE interaction}
There are many models of DM/DE interaction ( see \cite{uk} for a
review of interacting fluids). The thermodynamics of a
non-conserved energy-momentum in the cosmological models has been
discussed in \cite{lima4,jesus,pav}. We consider in this section a
particular class of models when the energy-momentum
non-conservation is expressed by the particle density.

 We restrict our discussion
to the metric
\begin{equation}
ds^{2}=g_{\mu\nu}dx^{\mu}dx^{\nu}=dt^{2}-a^{2}h_{jk}dx^{j}dx^{k}
\end{equation}(mainly to the FLWR metric when the tensor $h_{jk}$
describes the homogenous space of constant curvature). The
 energy-momentum tensor $\tilde {T}$ of the dark matter under general
assumptions can be expressed in the form
\begin{equation}
\tilde{T}^{\mu\nu}=(\tilde{\rho}+\tilde{p})u^{\mu}u^{\nu}-\tilde{p}g^{\mu\nu}+
\tilde{\Pi}^{\mu\nu},
\end{equation}
where   $\tilde{\Pi}^{\mu\nu}$ is the stress tensor and the
four-velocity $u^{\mu}$ satisfies the condition
\begin{equation}
g_{\mu\nu}u^{\mu}u^{\nu}=1.
\end{equation}In particular, the decomposition (16) holds true if the
dark matter is determined by a phase space distribution $\Omega$.
 In a homogeneous
universe described by the metric (15) the stress tensor
$\tilde{\Pi}^{\mu\nu}$ is absent.We introduce the energy momentum
tensor $T_{de}$ of an ideal fluid ( with zero stress tensor)
describing the dark energy
\begin{equation}
T^{\mu\nu}_{de}=(\rho_{de}+p_{de})u^{\mu}u^{\nu}-g^{\mu\nu}p_{de}.
\end{equation}
In general, we consider a total energy-momentum tensor which is
conserved and consists of many components. We assume the
energy-momentum (non)conservation law (with certain currents
$N_{b}^{\nu}$; the constant $\kappa$ is describing a dissipation)
\begin{equation}
(T^{\mu \nu}_{b})_{;\mu}=3\kappa^{2}N^{\nu}_{b}.
\end{equation}
We require
\begin{equation}
\sum_{b}N^{\nu}_{b}=0,
\end{equation} if the total energy-momentum
\begin{displaymath}
T^{\mu\nu}_{tot}=\sum_{b} T^{\mu\nu}_{b}
\end{displaymath}
is to be conserved.

We consider models such that $T_{tot}=\tilde{T}+T_{de}$ (so that the index $b$ involves only DM and DE) and
\begin{equation}
(\tilde{T}^{\mu\nu})_{;\mu;\nu}=3\kappa^{2}(\tilde{N}^{\nu})_{;\nu}=0
\end{equation}
This condition discriminates the model
\cite{habacqg,hss,habampl1}(introduced also in \cite{calo}) from
other models of interacting DM/DE systems where $N^{\nu}_{b}$ are
linearly related to the energy-momentum \cite{amendola,mar,sola}.
Eq.(21) in a homogeneous flat universe implies
\begin{equation}
\tilde{N}^{0}=a^{-3}\frac{\tilde{\gamma}}{3\kappa^{2}},
\end{equation}where we introduced another constant $\tilde{\gamma}$.
$\tilde{N}^{0}$ has an interpretation of the particle density
(as the particle density decays as an inverse of the volume). Eq.(19)
means that the increase of the energy density in an expanding
universe is proportional to the particle density.

 In
order to satisfy the total energy-momentum conservation (20) we
set
\begin{equation}
\gamma_{de}=-\tilde{\gamma}\equiv -\gamma.
\end{equation}

 Then, eqs.(19)-(23)  in the frame
$u=(1,{\bf 0})$ and in a homogeneous space give
\begin{equation}
\partial_{t}\rho_{de}+3H(1+w)\rho_{de}=-\gamma a^{-3},
\end{equation}where $H=a^{-1}\partial_{t}a$ and
\begin{displaymath}
w=\frac{p_{de}}{\rho_{de}}.\end{displaymath}
The solution of eq.(24) for a constant $w$ is
\begin{equation} \rho_{de}(t)=a^{-3(1+w)}\sigma_{de}(0)-\gamma
a(t)^{-3(1+w)}\int_{t_{0}}^{t}a(s)^{3w}ds\end{equation}
$\sigma_{de}(0) $ is a constant such that
\begin{displaymath}
\rho_{de}(t_{0})=a(t_{0})^{-3(1+w)}\sigma_{de}(0)
\end{displaymath}($\sigma_{de}(0) $ is just the initial DE density if we choose
the convention $a(t_{0})=1$). In the frame $u=(1,{\bf 0})$ eq.(19)
for $\tilde{T}^{\mu 0}$ reads
\begin{equation}
\partial_{t}\tilde{\rho}+3(1+\tilde{w})H\tilde{\rho}=\gamma  a^{-3}
\end{equation}
If we approximate $\tilde{w}$ as a constant then we get an analog
of eq.(25)
\begin{equation} \tilde{\rho}(t)=a^{-3(1+\tilde{w})}\tilde{\sigma}(0)+\gamma
a(t)^{-3(1+\tilde{w})}\int_{t_{0}}^{t}a(s)^{3\tilde{w}}ds.\end{equation}
For the relativistic diffusion of sec.4 $\tilde{w}$ is time
dependent $0<\tilde{w}(a)\leq\frac{1}{3}$. As discussed in the
next section $\tilde{w}(a)=\frac{1}{3}$ for massless particles.
For $\tilde{w}(a)=\frac{1}{3}$
\begin{displaymath} \tilde{\rho}(t)=a^{-4}\tilde{\sigma}(0)+\gamma
a(t)^{-4}\int_{t_{0}}^{t}a(s)ds.\end{displaymath}

In \cite{hss,ss} the energy densities (25) and (27) have been
inserted in Einstein equations which were solved numerically. The
expansion law $a(t)$ cannot be expressed by an elementary function.
Nevertheless, for a qualitative exposition let us insert as an
approximation  $a(t)=t^{\alpha}$. Then, with $ \tilde{w}>0$ the
diffusive term in eq.(27) (proportional to $\gamma$) is dominating
the expansion for a large time as $\tilde{\rho}\simeq \gamma
a^{-3+\frac{1}{\alpha}}$.  This domination  remains true for the
exponential expansion (appearing at large time in some of the solutions of
\cite{hss,ss}); if $a=\exp(Ht) $, then $\tilde{\rho}\simeq \gamma
a^{-3}$.

We apply the
 equilibrium  thermodynamic definition of the entropy (due to Clausius and Caratheodory)
  as the integrating factor for the heat $Q_{b}$
  to each component $b$ of
the cosmic fluid. Then, we have the Gibbs-Duhem differential (no
chemical potential)
\begin{equation}
TdS_{b}=dQ_{b}=dE_{b}+p_{b}dV=(\rho_{b}+p_{b})dV+Vd\rho_{b}
\end{equation}
with $pdV$ denoting the work done during the expansion of the
volume  $V=V_{0}\sqrt{g}$ where $g=\vert \det(g_{\mu\nu})\vert$ .
Using \begin{equation}
\partial_{\mu}\sqrt{g}=\Gamma^{\alpha}_{\alpha\mu}\sqrt{g},
\end{equation}where   $\Gamma$ is the Christoffel symbol,
from eqs.(26) and (28) in the frame moving with the fluid
($u=(1,0)$ ) we obtain for an ideal fluid of DM
\begin{equation}
Td\tilde{S}=(\partial_{\tau}\tilde{\rho}+\Gamma^{\alpha}_{\alpha
0}(\tilde{\rho}+\tilde{p}))V_{0}\sqrt{g}dt=
3\kappa^{2}\tilde{N^{0}}V_{0}\sqrt{g}dt.
\end{equation}Hence, from eq.(22) in the homogeneous flat universe

\begin{equation}
Td\tilde{S}=V_{0}\gamma dt.
\end{equation}
Let us compare now the formulae for the energy (7) and (27). We
can see that $B$ in eq.(7) cannot be a constant if there is an
energy non-conservation . In the derivation of $\rho$ in Sec.2 we
should take into account that $\rho$ depends on two variables $a$
and $A$ where

\begin{displaymath}A=\int_{t_{0}}^{t}a^{3w}ds.\end{displaymath} We can write $\tilde{\rho}$ (27) in the form
\begin{equation}
\tilde{\rho}=3ZB(\tilde{A})V_{0}^{-1}T_{0}^{-\frac{1+\tilde{w}}{\tilde{w}}}T_{a}^{\frac{1+\tilde{w}}{\tilde{w}}},
\end{equation}
where we denoted \begin{displaymath}
Z=\frac{\tilde{\sigma}(0)V_{0}}{3T_{0}}
\end{displaymath}and $\kappa^{2}=\frac{\gamma}{3Z}$
(we use this  complicated way of expressing the energy density
(27) in order to comply with the constants appearing in the
diffusion of Sec.4, Z is the statistical sum in the model of
sec.4) $T_{a}$ is the \textquoteleft temperature at fixed
$a$\textquoteright determined by eq.(9) and
\begin{equation} B(\tilde{A})=\frac{\tilde{\sigma}(0)V_{0}}{3Z}+\frac{\gamma V_{0}}{3Z}
\int_{t_{0}}^{t}a(s)^{3\tilde{w}}ds\equiv
T_{0}+\kappa^{2}\tilde{A}\end{equation} Next, we must modify the
formula for the entropy (8) derived through  an ``adiabatic''
variation of variables. We assume that the entropy $\tilde{S}$
consists of the ``adiabatic'' part $\tilde{S}_{ad}$ (8) and the
dissipative part $\tilde{S}_{\gamma}$ (31). Hence, taking into
account a varying $V$ as  in eqs.(3) and (5) and integrating
eq.(31) we obtain
\begin{equation}
\tilde{S}(t)=\tilde{S}_{ad}+\tilde{S}_{\gamma}=V(1+\tilde{w})\frac{\tilde{\rho}}{T}+V_{0}\gamma\int_{t_{0}}^{t}\frac{ds}{T(s)}\end{equation}
where $V=V_{0}a^{3}$.

If we demand that $\tilde{S}_{ad} $ does not change during the expansion then using the result (32)
we obtain the dependence of $T$ on $a$
\begin{equation} T=B(\tilde{A})T_{0}^{-1}T_{a}
\end{equation}
An insertion of eq.(35)   in eq.(34) gives
\begin{equation}
\tilde{S}=3Z\ln
B(\tilde{A})+V(1+\tilde{w})\frac{\tilde{\rho}}{T}=3Z\ln\Big(\frac{TT_{0}}{T_{a}}\Big)
+V(1+\tilde{w})\frac{\tilde{\rho}}{T}.\end{equation} Eq.(36)
agrees with the definition of temperature \cite{landau}
\begin{equation}
\frac{1}{T}=\frac{1}{V}\frac{\partial \tilde{
S}}{\partial\tilde{\rho}}
\end{equation}
In Sec.5 we calculate the entropy(here and in further discussions
we set the Boltzmann constant $k_{B}=c=\hbar=1$)
\begin{equation}
\tilde{S}=-\int d{\bf x}\frac{d{\bf p}}{(2\pi)^{3}}g\Omega
\ln\Omega
\end{equation}
for a system  of diffusing particles with a probability
distribution $\Omega({\bf x},{\bf p})$ in the phase space.
We show   that the formula (35) for temperature and the formula (32) for the energy density
are satisfied for this  realization of the dissipative system.

Writing an
 analog equation for $T_{de}dS_{de}$ we get eqs.(33)-(34) with $\gamma\rightarrow -\gamma$.
  The variation of the
    total entropy $\tilde{S}+S_{de}$ during universe expansion
    depends on the scale factor $a$ and will be discussed in Sec.6.
In the generalized second law of thermodynamics
\cite{bekenstein,gibbons,davis,cai,gar} the entropy of the
apparent horizon plays an important role contributing to the
entropy increase \cite{pavon} and compensating the entropy
decrease during expansion of other components. However, in models
of this section the entropy of the dark sector is increasing (see
Sec.6). So, the increase of the entropy of the horizon does not
play such a decisive role. In the Appendix we generalize the
derivations of this section to an arbitrary interaction
$\tilde{N}^{\nu}$ in eq.(19) (without an assumption of the current
conservation (21)). However, so far we have no  candidate for
$\tilde{N}^{\nu}$ other than the particle current.
\section{Relativistic diffusion}In this section we discuss a
Markovian approximation of an interaction of the system (DM) with
an environment (DE) which leads to a description of this
interaction by a diffusion. We consider a relativistic
generalization of the Krammers diffusion defined on the phase
space. It is determined in the unique way by the requirement that
the diffusing particle moves on the mass-shell $p^{2}=m^{2}$ (see
\cite{dudley,franchi,habapre,habacqg1}).

In the contravariant spatial coordinates $p^{j}$ on the mass shell
we define the Riemannian metric $G_{jk}$
\begin{displaymath}
ds^{2}=g_{\mu\nu}dp^{\mu}dp^{\nu}=-G_{jk}dp^{j}dp^{k},
\end{displaymath}
where $p_{0}$ is expressed by $p^{j}$.

The inverse matrix is  (we assumed that $ g_{0k}=0$)
\begin{equation}
G^{jk}=-g^{jk}m^{2}+p^{j}p^{k}.
\end{equation}
Then,
\begin{displaymath}
\Gamma\equiv \det(G_{jk})=\det(g_{jk})p_{0}^{-2},
\end{displaymath}
where
\begin{displaymath}
p_{0}^{2}=m^{2}-g_{jk}p^{j}p^{k}.
\end{displaymath}
 We define   diffusion as a stochastic process  generated by
the Laplace-Beltrami operator $\triangle_{H}^{m}$ on
 the mass shell
\begin{equation}
\triangle_{H}^{m}=\frac{1}{\sqrt{\Gamma}}\partial_{j}G^{jk}\sqrt{\Gamma}\partial_{k},
\end{equation}
where
 $\partial_{j}=\frac{\partial}{\partial
p^{ j}}$ .

The transport equation for the  diffusion generated by
$\triangle_{H}$ reads
\begin{equation}
\begin{array}{l}
(p^{\mu}\partial^{x}_{\mu}-\Gamma^{k}_{\mu\nu}p^{\mu}p^{\nu}\partial_{k})\Omega=
\kappa^{2}\triangle^{m}_{H}\Omega,
\end{array}\end{equation} where $\kappa^{2}$ is the diffusion constant and $\partial_{\mu}^{x}=\frac{\partial}{\partial
x^{ \mu}}$.

The relativistic diffusion determines the phase space
distribution. Then, we can define the current
\begin{equation}\begin{array}{l}\tilde{ N}^{\mu}=\sqrt{g}\int \frac{d{\bf
p}}{(2\pi)^{3}}p_{0}^{-1}p^{\mu}\Omega,
\end{array}
\end{equation}where $g=\vert\det( g_{\mu\nu})\vert$ and the energy-momentum

\begin{equation}\begin{array}{l}
\tilde{T}^{\mu\nu}=\sqrt{g}\int  \frac{d{\bf
p}}{(2\pi)^{3}}p_{0}^{-1}p^{\mu}p^{\nu}\Omega.
\end{array}
\end{equation} The form of the current (42) and the energy-momentum (43) follow from the Liouville
description of a stream of particles. Its tensor character is a
consequence of the invariance of the measure $\sqrt{g}d^{4}p
\delta(g_{\mu\nu}p^{\mu}p^{\nu}-m^{2})$ (where $d^{4}p$ is a
product of four infinitesimal coordinate translations) under
coordinate transformations \cite{wein,choquet}. It can be shown
for a general metric that the relation (19) is satisfied
\cite{habampl1}
\begin{equation}
(\tilde{T}^{\mu\nu})_{;\mu}=3\kappa^{2}\tilde{N}^{\nu},
\end{equation} where the current $\tilde{N}^{\mu}$ is conserved

\begin{equation}
(\tilde{N}^{\mu})_{;\mu}=g^{-\frac{1}{2}}\partial_{\mu}(g^{\frac{1}{2}}\tilde{N}^{\mu})=0.
\end{equation}
From eq.(45) we get eq.(22) in a homogeneous metric. If $\Omega$
is $x$-independent and $\int d{\bf x}=V_{0}$ then the constant
$\gamma$ can be expressed from eq.(42) (according to eq.(22)) as
\begin{equation}\frac{\gamma}{3\kappa^{2}}=\tilde{N}^{0}\sqrt{g}
=g\int  \frac{d{\bf p}}{(2\pi)^{3}}\Omega\equiv ZV_{0}^{-1},
\end{equation}
where $Z\equiv \int d{\bf x}g\int  \frac{d{\bf
p}}{(2\pi)^{3}}\Omega$ is the statistical sum.

 In the massless case from
$\tilde{T}^{\mu}_{\mu}=0$ we obtain $\tilde{w}=\frac{1}{3}$, then
from eq.(27)
\begin{equation}\begin{array}{l}
\tilde{\rho}(t)=a^{-4}\tilde{\sigma}(0)+\gamma
a(t)^{-4}\int_{t_{0}}^{t}a(s)ds\equiv
3ZV_{0}^{-1}a^{-4}(\frac{V_{0}}{3Z}\tilde{\sigma}(0)+\kappa^{2}\tilde{A})\cr=3ZV^{-1}a^{-4}(T_{0}+\kappa^{2}\tilde{A}),\end{array}\end{equation}
where $T_{0}\equiv \tilde{\sigma}(0)\kappa^{2}\gamma^{-1}=
V_{0}\tilde{\sigma}(0)(3Z)^{-1}$ .

We can solve  eq.(41) exactly when $m=0$ and the initial condition
is the J\"uttner distribution \cite{juttner}. The solution is
\cite{habacqg}
\begin{equation}\begin{array}{l}
\Omega^{J}(t,T_{0})=C_{R}T_{0}^{3}(T_{0}+\kappa^{2}\tilde{A})^{-3}\exp\Big(-\frac{a^{2}}{T_{0}+\kappa^{2}\tilde{A}}\vert{\bf
p}\vert\Big),\end{array}
\end{equation}
where $T_{0}$ is a free parameter such that $T_{0} a(t_{0})^{-1} $
can be interpreted as a temperature at $t=t_{0}$ (we assume $
a(t_{0})=1$; from eq.(47) $T_{0}$ is related to
$\tilde{\sigma}(0)$ and to $Z$). We introduced a constant $C_{R}$
which determines the normalization of $\Omega^{J}$.  If we
normalize $\Omega^{J}$ to the particle density $n$ at $t_{0}$
\begin{equation}
\int d{\bf
p}\frac{1}{(2\pi)^{3}}g\Omega^{J}(T_{0})=n\equiv\frac{N}{V_{0}},
\end{equation}where $N$ is the number of particles, then
\begin{equation}
C_{R}=n\pi^{2}\frac{ 1}{T_{0}^{3}}.\end{equation} Summing over the
temperatures $T_{0}$ with  a density $q(T_{0})$
\begin{equation}
\Omega(t)=\int dT_{0}q(T_{0})\Omega^{J}(T_{0},t)
\end{equation}
we get a more general solution of the diffusion equation.

In \cite{habacqg1} we have derived an exact solution of the
diffusion equation (41) with $m\geq 0$ in de Sitter space when
$a(t)=\exp(H(t-t_{0}))$. Then, the solution is
\begin{equation}
\Omega(t)=C_{R}\exp(-3Ht)\exp\Big(-\frac{H}{\kappa^{2}}\sqrt{a^{2}{\bf
p}^{2}+m^{2}}\Big).
\end{equation}
In \cite{haba16} we have shown that if $a(t)$ is close to an
exponential function then we can obtain an expansion around the
solution (52).

 We can calculate now explicitly from eq.(48)
\begin{equation}\begin{array}{l}
\tilde{N}^{0}= C_{R}\frac{1}{(2\pi)^{3}}8\pi
T_{0}^{3}a^{-3}.\end{array}
\end{equation}
From
\begin{displaymath}\partial_{\tau}\tilde{T}^{00}+4H\tilde{T}^{00}=\gamma
a^{-3}
\end{displaymath}
and eq.(47) we obtain
 \begin{equation}
 \gamma=\frac{1}{(2\pi)^{3}}24\pi\kappa^{2}T_{0}^{3}=3\kappa^{2}Z_{J}V_{0}^{-1}=3\kappa^{2}n.
\end{equation}
It is still interesting to consider the non-relativistic limit of
eq.(41)\cite{habacqg,hss}
\begin{equation}
m^{2}a^{2}\rightarrow \infty.
\end{equation}
Then, the limit of eq.(41) is
\begin{equation}
m^{-1}\kappa^{-2}(\partial_{t}-2Hp^{j}\partial_{j})\Omega=a^{-2}\triangle_{\bf
p}\Omega,
\end{equation}
where $\triangle_{\bf p}$ is the Laplacian in the momentum space.
This is the non-relativistic diffusion equation in an expanding
momentum space (see \cite{diff}). A special solution (with a
Maxwell-Boltzmann initial condition) of the diffusion equation
(56) reads
\begin{equation} \Omega_{NR}(t,T_{0},{\bf p},{\bf k})=C_{NR}T_{0}^{\frac{3}{2}}
(T_{0}+\kappa^{2}A_{NR})^{-\frac{3}{2}}\exp \Big(-\frac{(a^{2}{\bf
p}-a^{2}{\bf k})^{2}}{2m(T_{0}+\kappa^{2}A_{NR})}\Big),
\end{equation} where
 \begin{displaymath}
 C_{NR}=nT_{0}^{-\frac{3}{2}}(2\pi m)^{\frac{3}{2}}
 \end{displaymath}
which follows from the normalization
\begin{displaymath}
Z=\int  \frac{d{\bf p}}{(2\pi)^{3}}g\Omega=n
\end{displaymath}and\begin{equation}
A_{NR}=2\int_{0}^{t}a^{2}(s)ds.
\end{equation}
The  solution with a general initial condition is
\begin{displaymath}
\Omega_{t}({\bf p})=\int d{\bf k}dT_{0} \Omega_{NR}(t,T_{0},{\bf
p},{\bf k})q(T_{0},{\bf k}).
\end{displaymath}

\section{Entropy, free energy  and temperature of the diffusive dark matter}
We have discussed the first law of equilibrium thermodynamics in
sec.2. The expansion of the universe can hardly be considered as a
sequence of infinitesimal transitions between equilibrium states.
For an expanding universe we need a formulation of non-equilibrium
thermodynamics.
 If a phase space evolution $\Omega$ is given then we can define
 the relative entropy for any two solutions $\Omega$ and
 $\Omega_{J}$ of the diffusion equation
 (41)\cite{lebowitz,risken}
 \begin{equation}
 {\cal F}=Z^{-1}\frac{1}{(2\pi)^{3}}\int d{\bf x} d{\bf
 p}g\Omega\ln\Big(Z^{-1}Z_{J}\Omega\Omega_{J}^{-1}\Big),
 \end{equation}
where
\begin{equation}
Z=\frac{1}{(2\pi)^{3}}\int  d{\bf x} d{\bf p}g\Omega,
\end{equation}\begin{equation}
Z_{J}=\frac{1}{(2\pi)^{3}}\int d{\bf x}d{\bf p}g\Omega_{J}.
\end{equation}
 It can be shown that owing to the conservation of the
current $N^{\mu}$ the normalizing integrals  $Z$ and $Z_{J}$ do
not depend on time. We have (the proofs follow ref.\cite{risken};
a slight extension is needed taking into account the time
dependence of the metric)
\begin{equation}
{\cal F}(\Omega)\geq 0
\end{equation}
${\cal F}(\Omega_{J})=0$ and \begin{equation}
\partial_{t}{\cal F}=-Z^{-1}\frac{1}{(2\pi)^{3}}\int  d{\bf x}d{\bf p}g\Omega G^{jk}\partial_{j}\ln Q
\partial_{k}\ln Q\leq 0\end{equation}where
\begin{equation}
Q=\Omega\Omega_{J}^{-1}.
\end{equation}
Let us consider in more detail the limit  $ a^{2}m^{2}\rightarrow
0$ in the diffusion equation (41). We  choose  the solution
$\Omega^{J}(T_{0})$ (48) as $\Omega^{J}$ in ${\cal F}$. Define
\begin{equation}\Phi=Z{\cal F}
-Z\ln(Z_{J}Z^{-1})-3Z\ln\Big(C_{R}^{-\frac{1}{3}}T_{0}^{-1}(T_{0}+\kappa^{2}\tilde{A})\Big).
\end{equation}
Then, \begin{displaymath} \partial_{t}\Phi \leq 0\end{displaymath}
Consider
\begin{equation}U=
\frac{a\tilde{E}}{(T_{0}+\kappa^{2}\tilde{A})}\equiv\frac{\tilde{E}}{T},
\end{equation}where\begin{equation} \tilde{E}=\int d{\bf x}\sqrt{g}\tilde{T}^{00}=\frac{1}{(2\pi)^{3}}\int d{\bf
p}d{\bf x}p^{0}g\Omega=\int d{\bf x}\sqrt{g}\tilde{\rho}.
\end{equation}
 It follows from eq.(65) and (59) that $\tilde{F}=\Phi T$ satisfies
 the thermodynamic identity defining the free energy $\tilde{F}$
\begin{equation}\frac{\tilde{F}}{T}
=-\tilde{S}+U=-\tilde{S}+\frac{1}{T}\tilde{E},
\end{equation}where $\tilde{S}$ is defined in eq.(38) and \begin{equation}
T(t)=\frac{T(t_{0})+\kappa^{2}\tilde{A}}{a}.
\end{equation}The formula (69) for the temperature results from eqs.(37)and (68) through
the basic principles of statistical thermodynamics \cite{landau}
(it also follows from the J\"uttner distribution
(48))\begin{equation} \frac{1}{T(t)}=\frac{\partial
\tilde{S}}{\partial \tilde{E}}.
\end{equation}
The  evolution law $a(t)$ depends on the solution of Einstein
equations discussed in \cite{hss,ss}. As an example, for a
power-law evolution the temperature (69) is increasing and
$T(t)\rightarrow \infty$ as $t\rightarrow \infty$. However, the
power-law evolution is considered in cosmology  only for an
intermediate time. For a large time we expect constant
acceleration $a(t)=\exp(Ht)$ resulting from the cosmological
constant. Then, the temperature tends to a finite value
\begin{equation}
T(t)\rightarrow \kappa^{2}H^{-1}.
\end{equation}
 We could calculate this temperature inserting
the formula (52) for $\Omega$ in de Sitter space into the formula
(38) for the entropy and calculate $T$ from eq.(70).

Let us calculate the entropy of $\Omega^{J}$ from eq.(38)
\begin{equation}
\tilde{S}^{J}=3nV_{0}-nV_{0}\Big(
\ln(n\pi^{2})-3\ln(T_{0}+\kappa^{2}A)\Big).
\end{equation}
This formula coincides (up to a constant) with eq.(34) derived on
the basis of the first law  of thermodynamics. The dependence of
entropy on temperature (70) coincides with the standard one for
J\"uttner distribution (although now the temperature depends on
time according to eq.(69)).

 We can repeat the formulation of the
non-equilibrium thermodynamics in the non-relativistic limit
$a^{2}m^{2}\rightarrow \infty$ defining
\begin{equation}{\cal F}(\Omega)=(2\pi)^{-3}Z^{-1}\int d{\bf x}d{\bf p}g \Omega\ln(Z^{-1}\Omega Z_{NR}\Omega_{NR}^{-1})
\end{equation} where $\Omega_{NR}$ denotes the Maxwell-Boltzmann distribution (57)
with ${\bf k}=0$ . We have ${\cal F}(\Omega_{NR})=0$,${\cal F}\geq
0$ and  $\partial_{t}{\cal F}\leq 0$. We define
\begin{equation}\Phi
=Z{\cal
F}-Z\ln(Z_{NR}Z^{-1})-\frac{3}{2}Z\ln\Big(C_{NR}T_{0}^{-1}(T_{0}+\kappa^{2}A_{NR})\Big)
\end{equation}with
\begin{displaymath}\begin{array}{l} Z_{NR}=(2\pi)^{-3}\int d{\bf
p}d{\bf x}g\Omega_{NR}=C_{NR}T_{0}^{\frac{3}{2}}
(2\pi)^{-3}(2m\pi)^{\frac{3}{2}}\end{array}
\end{displaymath}where
\begin{displaymath}
C_{NR}=nT_{0}^{-\frac{3}{2}}(2\pi)^{\frac{3}{2}}m^{\frac{3}{2}}
\end{displaymath}
We have from the definition of ${\cal F}$ and the free energy
$\tilde{F}=\Phi T$
\begin{equation}\frac{\tilde{F}}{T}
=-\tilde{S}+\frac{\tilde{E}_{nr}}{T}
\end{equation}
where\begin{equation}
\frac{\tilde{E}_{nr}}{T}=\frac{a^{5}V_{0}}{T_{0}+\kappa^{2}A_{NR}}\tilde{\rho}_{nr}=\frac{V}{T}\tilde{\rho}_{nr}
\end{equation} and
\begin{equation}
\tilde{\rho}_{nr}=\sqrt{g}\int d{\bf p}\frac{{\bf
p}^{2}a^{2}}{2m}\Omega_{NR}
\end{equation}
with the temperature  defined by
\begin{equation}
\frac{1}{T}=\frac{\partial \tilde{S}}{\partial
\tilde{E}_{nr}}=\frac{a^{2}}{T_{0}+\kappa^{2}A_{NR}}.\end{equation}
The $a^{-2} $ decay of the temperature in the non-relativistic non-diffusive case
$\kappa^{2}=0$ is well known \cite{bernstein}.The
calculation of entropy from the Gibbs formula (38) gives for the
Maxwell-Boltzmann temperature state (57) the result
\begin{equation}
S(\Omega_{NR})=\frac{3}{2}Z_{NR}+\frac{3}{2}Z_{NR}\ln\Big(C_{NR}^{-\frac{2}{3}}T_{0}^{-1}(T_{0}+\kappa^{2}A_{NR})\Big)
+V\frac{\rho_{nr}}{T},\end{equation} which up to a constant agrees
with the result (34)(with $\tilde{w}=0$).Its dependence on
temperature coincides with the standard formulas of classical
statistical physics of ideal gases.
\section{Evolution of the temperature and entropy}
In this section we study the entropy and  temperature evolution
law in more detail. For a general solution $a(t)$ of Einstein
equations   the temperature (35) is an involved function of time.
We have
\begin{equation}
T(t)=a^{-3\tilde{w}}\Big(T_{0}+\kappa^{2}\int_{t_{0}}^{t}a^{3\tilde{w}}(s)ds\Big)
\end{equation}and \begin{displaymath}
\partial_{t}T=\kappa^{2}-3\tilde{w}H T.
\end{displaymath}
For $\tilde{w}>0$ ( DM) and large $t$ the temperature is an
increasing function of time because
$\int_{t_{0}}^{t}a^{3\tilde{w}}>> a^{3\tilde{w}}$ for $t>>t_{0}$.

 From
eq.(35) for DE ( $\gamma\rightarrow -\gamma$) we obtain the
formula
\begin{equation}
T_{de}=a^{3\vert w\vert}\Big(T_{de
0}-\kappa^{2}\int_{t_{0}}^{t}dsa^{3w}\Big).
\end{equation}
Hence,
\begin{equation}
\partial_{t}T_{de}=-\kappa^{2}-3wH T_{de}.
\end{equation}
 In order to be more specific on the time dependence we must  know the form of $a(t)$.
 In the diffusion model some numerical results on the evolution $a(t)$ have been obtained
 in \cite{hss,ss}. However, a part of the discussion of the
 entropy and temperature does not depend on the details of
 Einstein equations but rather on the dissipation law (as can be
 seen from the discussion in the Appendix). For this reason it seems relevant
to admit various possible approximate evolutions $a(t)$ in order to discuss their 
consequences for temperature and entropy.
  If we take, as an approximation of the evolution law for a large time , $a=\exp(H(t-t_{0}))$ then, if
$T_{0}>T_{\infty}\equiv \frac{\kappa^{2}}{3H\tilde{w}}$ then the
temperature of DM is monotonically decreasing to its minimal value
$T_{\infty}=\frac{\kappa^{2}}{3\tilde{w}H}$. If $T_{0}<T_{\infty}$
then the temperature of DM is monotonically increasing to its
maximal value $\frac{\kappa^{2}}{3\tilde{w}H}$. For DE, if $T_{de
0}>\frac{\kappa^{2}}{3H\vert w\vert}$ then for the exponential
expansion, the temperature is exponentially increasing to
infinity. If $T_{de 0}<\frac{\kappa^{2}}{3H\vert w\vert}$ then the
DE temperature becomes negative and the entropy does not make
sense. For a power-law expansion $a(t)=t_{0}^{-\alpha}t^{\alpha}$
the temperature of DM grows linearly. The temperature of DE is
non-negative for all $t\geq t_{0}>0$  only if  $3\alpha w+1<0$ and
$ T_{0}\geq\frac{\kappa^{2}}{\vert 3w\alpha+1\vert}t_{0}$. Then,
$\int_{t_{0}}^{\infty}ds a^{3w}<\infty$ in eq.(81). If $3\alpha
w+1>0$ then the temperature becomes negative for a large time and
the entropy does not make sense. For an exact description of the
temperature evolution we need numerical methods for each
trajectory $a(t)$.

The entropy of the DM/DE system is a sum of entropies.  We do not
know a phase space distribution of DE. However, from the energy
conservation (eqs.(19)-(20) and (31) ) it follows that
\begin{equation}
\partial_{t}S=\partial_{t}\tilde{S}+\partial_{t}S_{de}=V_{0}\gamma
\Big(\frac{1}{T(s)}-\frac{1}{T_{de}(s)}\Big).
\end{equation}Hence,
\begin{displaymath}
\partial_{t}S\geq 0
\end{displaymath}
if $T\leq T_{de} $.
 This inequality  is certainly satisfied for an exponential expansion
  (as long as the temperature
makes sense) because we  have the inequality
\begin{equation}
T_{de}\geq a^{3\vert w\vert}(T_{de
0}-\kappa^{2}\int_{t_{0}}^{\infty}a^{3w}(s)ds)
\end{equation} and for an exponential expansion $T(t)$
is bounded. In general, we have $T(t)\leq Ct$ for a large time
(with a certain constant $C>0)$. For a power law expansion we have
additionally $ C_{0}t \leq T(t)\leq Ct$ with a certain $C_{0}>0$.
Then, a sufficient condition for $T\leq T_{de}$ ensuring an
increase of the  entropy at large time is satisfied if
\begin{equation}
a^{3\vert w\vert}> K t
\end{equation} with a sufficiently large $K$.
For a power-law expansion this means $3\alpha \vert w\vert>1$; the
condition which is also necessary  for the temperature to be
non-negative for all time. Note that the conditions
$w<-\frac{1}{3}$ and   $\alpha>1$ are also needed for an
accelerated expansion.

As in eq.(34) we can integrate eq.(83)
\begin{equation}
S=\tilde{S}+S_{de}=V_{0}\gamma\int_{0}^{t} ds
\Big(\frac{1}{T(s)}-\frac{1}{T_{de}(s)}\Big)
+V(1+\tilde{w})\frac{\tilde{\rho}}{T}+V(1+w)\frac{\rho_{de}}{T_{de}}\end{equation}
Using eqs.(80)-(81) (or returning to eq.(35) ) we obtain
\begin{equation}\begin{array}{l} S=3Z\ln\Big(T_{0}+\kappa^{2}\int_{t_{0}}^{t}
ds a(s)^{3\tilde{w}}\Big)+3Z \ln\Big(T_{de
0}-\kappa^{2}\int_{t_{0}}^{t} ds
a(s)^{3w}\Big)\cr+V(1+\tilde{w})\frac{\tilde{\rho}}{T}+V(1+w)\frac{\rho_{de}}{T_{de}}.
\end{array}\end{equation}
The first (logarithmic) term is increasing as a function of time
whereas the second term is decreasing.  If
\begin{equation}
\int_{t_{0}}^{\infty}ds a^{3w}<\infty
\end{equation}then the second term  has a lower bound $\ln\Big(T_{de
0}-\kappa^{2}\int_{0}^{\infty} ds a(s)^{3w}\Big)$. The condition
(88)is necessary for a definition of entropy of DE as the argument
of the logarithm must be positive.  As discussed below eq.(82)
 the argument of the logarithm is negative for negative temperatures
 as can happen for DE with a power-law expansion such that $3\alpha w>-1$.
 For systems which have a meaningful entropy of the dark sector the entropy
 increase is contained in the  first term of eq.(87). The terms depending on
the energy density in eq.(87) are constant as follows from the
derivation of the temperature dependence in sec.2. It follows from
eq.(83) that $\partial_{t}S>0$ for $T<T_{de}$. As discussed in
\cite{pavon} the total entropy should tend to a maximum what
requires $\partial_{t}^{2}S_{tot}< 0$ where $S_{tot}$ in
\cite{pavon} includes the entropy of the apparent horizon which
has a negative second order derivative for some range of $w_{eff}$
(where $w_{eff}$ is the effective equation of state). In the case
of the dark sector (86) we have from eqs.(80)-(82)
\begin{equation}
\partial_{t}^{2}S=-\gamma\kappa^{2}(T^{-2}+T_{de}^{-2})
+3\gamma\tilde{w}\frac{H(t)}{T}-3\gamma w\frac{H(t)}{T_{de}}.
\end{equation}
The first term is negative whereas the remaining terms are
positive. We need detailed calculations to estimate the balance of
these terms. We have calculated  these expressions for the
exponential expansion $a=\exp(H(t-t_{0}))$ then
\begin{equation}
\partial_{t}^{2}\tilde{S}=3\tilde{w}H
T^{-2}V_{0}\gamma(T_{0}-\frac{\kappa^{2}}{3\tilde{w}H})\exp(-3\tilde{w}H(t-t_{0}))
\end{equation}
and
\begin{equation}
\partial_{t}^{2}S_{de}=3\vert w\vert H
T_{de}^{-2}V_{0}\gamma(T_{0}-\frac{\kappa^{2}}{3\vert w\vert
H})\exp(3\vert w\vert(t-t_{0})).
\end{equation}
For a large time
\begin{equation}
\partial_{t}^{2}\tilde{S}\simeq (3\tilde{w}H)^{3}\kappa^{-4}
V_{0}\gamma(T_{0}-\frac{\kappa^{2}}{3\tilde{w}H})\exp(-3\tilde{w}H(t-t_{0}))
\end{equation}
and
\begin{equation}
\partial_{t}^{2}S_{de}\simeq 3\vert w\vert H
V_{0}\gamma\Big(T_{0}-\frac{\kappa^{2}}{3\vert w\vert
H}\Big)^{-1}\exp(-3\vert w\vert H(t-t_{0})).\end{equation}
$\partial_{t}^{2}S_{de}>0$ because the dominator in eq.(93) must
be positive if the temperature is to be positive. Then, the second
time derivative of the entropy $S$ can be negative only if $\vert
w\vert > \tilde{w}$, so that the second derivative of
$S_{\Lambda}$ is decaying faster than the one of $\tilde{S}$.
Moreover, for $\partial^{2}_{t}S<0$ we need
\begin{equation}
T_{0}<T_{\infty}=\frac{\kappa^{2}}{3\tilde{w}H}
\end{equation}
(then the temperature $T$ is increasing to its maximal value
$T_{\infty}$). For power-law expansion
$a(t)=t_{0}^{-\alpha}t^{\alpha}$ we can show that (under the
condition $3\vert w\vert\alpha>1$ of non-negative temperature)
$\partial_{t}^{2}S_{de}$ is decaying faster than
$\partial_{t}^{2}\tilde{S}$. Hence, the latter is dominating at
large time with the result
\begin{equation}
\partial^{2}_{t}S\simeq -\gamma
V_{0}\kappa^{-2}(3\tilde{w}+1)t^{-2}<0.
\end{equation}
The  entropy slowly achieves its maximum at an infinite time.

\section{Summary}It is known that negative
pressure leads to a curious behaviour of thermodynamic functions
of ideal non-interacting fluids. If we extend the argument to an
expanding universe then we obtain a non-conventional dependence of
temperature on the expansion scale factor $a$. In particular, the
temperature of DE is increasing as a function of $a$. In this
paper we assumed that the cosmological fluid consists of
components which can exchange the energy. We have suggested a
particular form of the energy (non) conservation assuming that the
energy gain (or loss) is proportional to the particle density. We
have  obtained this energy exchange in a model of a relativistic
diffusion of DM particles in a DE environment. We have shown that
as a result of DM/DE interaction the dependence of the temperature
on the scale factor undergoes a substantial change. The
temperature of DM (which has positive pressure, $\tilde{w}>0$) is
growing to infinity as a function of time for power law expansion.
For an exponential expansion it is reaching a fixed value
$\frac{\kappa^{2}}{3\tilde{w}H}$ as time tends to infinity. The
temperature of DE is increasing in all cases when the definition
of temperature  makes sense for arbitrarily large time. The
entropy is increasing for a large time because the temperature of
DM is increasing at a slower rate than the temperature of DE. This
is a consequence of the energy transfer from DE to DM. We derive
formulae for temperature and entropy in the case of more general
DM/DE interactions. However, we do not have  realizations of other
interactions besides the one of the relativistic diffusion.  The
study of temperature, entropy and the heat flow in various
components of the cosmological fluid should give important
information about structure formation and its evolution.

{\bf Acknowledgement}

The research is supported by the NCN grant DEC-2013/09/BST2/03455.
The author thanks an anonymous referee for  suggestions of an
improvement of the original manuscript.

\section{Appendix}
In the appendix we consider a more general system of DM/DE
interactions. Let us denote $3\kappa^{2}N^{0}=\gamma Q$ in
eq.(19). Then, in eq.(27)

\begin{equation} \tilde{\rho}(t)=a^{-3(1+\tilde{w})}\tilde{\sigma}(0)+\gamma
a(t)^{-3(1+\tilde{w})}\int_{t_{0}}^{t}a(s)^{3(1+\tilde{w})}Q(s)ds.\end{equation}
We write $\tilde{\rho}$ again in the form (32) with

\begin{equation} B(\tilde{A})=\frac{\tilde{\sigma}(0)V_{0}}{3Z}+\frac{\gamma V_{0}}{3Z}
\int_{t_{0}}^{t}a(s)^{3(1+\tilde{w})}Q(s)ds\equiv
T_{0}+\kappa^{2}\tilde{A},\end{equation} where $\kappa^{2}=
\frac{\gamma V_{0}}{3Z}$ and
\begin{displaymath}
\tilde{A}=\int_{t_{0}}^{t}a(s)^{3(1+\tilde{w})}Q(s)ds.
\end{displaymath}
Demanding that $S_{ad}$ in eq.(34) does not depend on time and
using eq.(96) we derive the evolution of temperature
\begin{equation}
T(t)=a(t)^{-3\tilde{w}}\Big(T_{0}+\frac{\gamma V_{0}}{3Z}
\int_{t_{0}}^{t}a(s)^{3(1+\tilde{w})}Q(s)ds\Big).
\end{equation}
The formulae for DE are obtained from eqs.(96)-(98) by
$\gamma\rightarrow-\gamma$ and $\tilde{w}\rightarrow w$. The
entropy is
\begin{equation}\begin{array}{l}S=\tilde{S}+S_{de}=
3Z\ln\Big(T_{0}+\kappa^{2}\tilde{A}\Big) +3Z \ln\Big(T_{de
0}-\kappa^{2}A\Big)\cr+V(1+\tilde{w})\frac{\tilde{\rho}}{T}+V(1+w)\frac{\rho_{de}}{T_{de}}
\end{array}\end{equation} where\begin{displaymath}
A=\int_{t_{0}}^{t}a(s)^{3(1+w)}Q(s)ds.
\end{displaymath}The last two terms  in eq.(99) are time
independent. As can be seen from Sec.6 the conclusions concerning
the growth of temperature and entropy depend on the sign of
$\gamma$ and on whether $Q$ grows or decays as a function of $a$.

\end{document}